\documentclass{svmult}

\usepackage{makeidx}         
\usepackage{graphicx}        
\usepackage{multicol}        
\usepackage[bottom]{footmisc}

\makeindex             

\begin{document}

\title*{Comprehensive Analysis of Dewetting Profiles to Quantify Hydrodynamic Slip}
\titlerunning{Analysis of Dewetting Profiles}
\author{Oliver B\"aumchen,\inst{1}\
Renate Fetzer,\inst{1,2}\ Andreas M\"unch,\inst{3}\ Barbara
Wagner,\inst{4}\ and Karin Jacobs\inst{1}}
\authorrunning{O. B\"aumchen, R. Fetzer, A. M\"unch, B. Wagner, and K.
Jacobs}
\institute{Department of Experimental Physics, Saarland University,
D-66123 Saarbruecken, Germany.
\texttt{o.baeumchen@physik.uni-saarland.de} \and Present address:
Ian Wark Research Institute, University of South Australia, Mawson
Lakes SA 5095, Australia. \and School of Mathematical Sciences,
University of Nottingham, University Park, Nottingham, NG7 2RD, UK.
\and Weierstrass Institute for Applied Analysis and Stochastics
(WIAS), Mohrenstrasse 19, D-10117 Berlin, Germany.}

%
%

\maketitle

Hydrodynamic slip of Newtonian liquids is a new phenomenon, the
origin of which is not yet clarified. There are various direct and
indirect techniques to measure slippage. Here we describe a method
to characterize the influence of slippage on the shape of rims
surrounding growing holes in thin polymer films. Atomic force
microscopy is used to study the shape of the rim; by analyzing its
profile and applying an appropriate lubrication model we are able
to determine the slip length for polystyrene films. In the
experiments we study polymer films below the entanglement length
that dewet from hydrophobized (silanized) surfaces. We show that
the slip length at the solid/liquid interface increases with
increasing viscosity. The correlation between viscosity and slip
length is dependent on the type of silanization. This indicates a
link between the molecular mechanism of the interaction of polymer
chains and silane molecules under flow conditions that we will
discuss in detail.

\section{Introduction}
\label{sec:1}


\subsection{Slippage at Solid/Liquid Interfaces}
\label{sec:1.1} The control of the flow properties at the
solid/liquid interface is important for applications ranging from
microfluidics, lab-on-chip devices to polymer melt extrusion.
Slippage would, for instance, greatly enhance the throughput in
lab-on-chip devices and extruders. Usually slippage is
characterized by the so-called slip length which is defined as the
distance between the solid/liquid interface and the point within
the solid where the velocity profile of the liquid extrapolates to
zero. During the last years, several different techniques have
been established to probe the slip length of different systems.
These methods can be classified into two groups, direct and
indirect measurements of the flow velocity at the solid/liquid
interface. Direct measurements are based on particle imaging
velocimetry techniques \cite{Tre02,Tre04,Lum03} that utilize
tracer particles or fluorescence recovery after photobleaching
\cite{Pit00,Sch05}. Colloidal probe atomic force microscopy
\cite{Cra01,Vin03} and surface forces apparatuses
\cite{Cot02,Zhu02} are more indirect techniques to probe slippage.
Detailed information concerning slippage and techniques to measure
liquid velocities in the vicinity of solid/liquid interfaces are
reviewed in recent articles by Neto et al. \cite{Net05}, Lauga et
al. \cite{Lau05} and Boquet and Barrat \cite{Boc07}.

\subsection{Dewetting Dynamics of Polymer Films}
\label{sec:1.2} Dewetting takes place whenever a liquid layer can
reduce energy by retracting from the contacting solid \cite{Ruc74}.
Dewetting starts by the birth of holes, which can be generated by
three different mechanisms, spinodal dewetting, homogeneous and
heterogeneous nucleation \cite{See012}. The holes grow in size, c.f.
Fig. \ref{fig:holes}, until neighboring holes touch. The quasi-final
state is a network of droplets \cite{Rei92}. The equilibrium state
would be a single droplet, yet this state is usually not awaited
since it may take years for viscous liquids such as our polystyrene.
Since we are only interested in the growth of holes and their rim
morphology, the underlying mechanism of their generation is
irrelevant. However, the most likely process in our system is
nucleation. The type of nucleus is unclear and also of no relevance
for our studies. It can be a dust particle or a heterogeneity in the
film or on the substrate. After holes are generated, they instantly
start to grow until they touch neighboring holes and coalesce. We
study the flow dynamics of thin polymer films on smooth hydrophobic
substrates. The driving force for the dewetting process can be
characterized by the spreading parameter $S$, which depends on the
surface tension of the liquid $\gamma_{lv}$ ($30.8\,$mN/m for
polystyrene) and the Young's contact angle $\Theta$ of the liquid on
top of the solid surface:

\begin{equation}\label{Spreitungsparameter}
  S=\gamma_{lv} (cos\Theta-1)
\end{equation}

The system also dissipates energy, namely by viscous friction
within the liquid and sliding friction at the solid/liquid
interface. A force balance between driving forces and dissipation
determines the dewetting rate. Conservation of mass leads to a rim
that surrounds each hole. We show that the shape of this rim is
not only sensitive to the chain length of the polymer melt
\cite{See011} but is also to the underlying substrate. The use of
recently developed models \cite{Fet071} enables us to extract the
slip length $b$ as well as the capillary number $Ca$ from the rim
profiles. The latter is given by

\begin{equation}\label{Capillary number}
  Ca=\frac{\eta \dot{s}}{\gamma_{lv}}
\end{equation}

Here, $\dot{s}$ is the current dewetting velocity, which can be
obtained by a series of optical images of the hole growth before the
sample is quenched to room temperature and imaged by atomic force
microscopy (AFM). From the capillary number, the viscosity $\eta$ of
the melt can be calculated and compared to independent viscosity
measurements, cf. Sect. \ref{sec:3.4}.



%
%
%
%
%
%
%
%
%
%
%
%
%
%

\section{Experimental Section}
\label{sec:2}

\subsection{Our System}
\label{sec:2.1} Atactic polystyrene (PS) obtained from PSS Mainz
with a molecular weight of 13.7 kg/mol ($M_\mathrm{w}/M_\mathrm{n} =
1.03$) was used as a liquid in our experiments. As substrates, we
used Si wafers (Siltronic AG, Burghausen, Germany) that were
hydrophobized by two different types of silanes following standard
methods \cite{Was89}. Thin PS films were prepared by spin coating a
toluene solution of PS onto mica, floating the films onto Millipore
water, and then picking them up with the silanized silicon wafers.
The floating step is necessary since on the spin coater, a drop of
toluene solution would just roll off the hydrophobized surface. All
PS films in this study have a thickness of $130(5)\,$nm. We utilized
two different silane coatings on the Si wafer (2.1 nm native oxide
layer): octadecyltrichlorosilane (OTS) and the shorter
dodecyltrichlorosilane (DTS), respectively. The thicknesses of these
self-assembled monolayers are $d_{\mathrm{OTS}}=2.3(2)\,$nm and
$d_{\mathrm{DTS}}=1.5(2)\,$nm as determined by ellipsometry (EP$^3$
by Nanofilm, Goettingen, Germany). The contact angle hysteresis of
water is very low ($6^{\circ}$ in case of OTS and $5^{\circ}$ in
case of DTS), while the advancing contact angles are
$116(1)^{\circ}$ (OTS) and $114(1)^{\circ}$ (DTS). Surface
characterization by AFM (Multimode by Veeco, Santa Barbara, CA, USA)
revealed RMS roughnesses of $0.09(1)\,$nm (OTS) and $0.13(2)\,$nm
(DTS) on a 1\,$\umu$m$^2$ area, and an (static) receding contact
angle of polystyrene droplets of $67(3)^{\circ}$ on both substrates.

Surface energies cannot be determined directly from contact angle
measurements of PS only, due to the fact that polystyrene has polar
contributions. The advancing contact angles of apolar liquids like
bicyclohexane vary slightly on both coatings. We find
$45(3)^{\circ}$ on OTS and $38(4)^{\circ}$ on DTS. The surface
energy of the substrate is linked to the contact angle of apolar
liquids via the Good-Girifalco equation \cite{Goo60}. Consequently,
we find a slightly larger surface energy for DTS
($\gamma_{\mathrm{DTS}}=26.4\,$mN/m) than for OTS
($\gamma_{\mathrm{OTS}}=23.9\,$mN/m) substrates. Identical contact
angles of PS on both substrates therefore lead to different energies
at the OTS/PS and DTS/PS interface due to Young's equation.

\subsection{Hole Growth Dynamics}
\label{sec:2.2} 
\begin{figure}
\centering
\includegraphics[height=3cm]{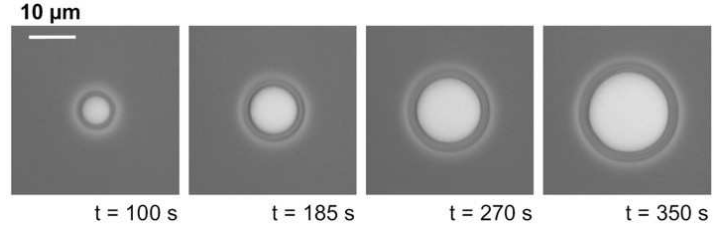}
\caption{Series of optical images of a growing hole in a PS film on
DTS at 120$^{\circ}$C.}
\label{fig:holes}       
\end{figure}

To induce dewetting, the films were heated to
different temperatures above the glass transition temperature of the
polymer. After a short time circular holes appear and instantly
start to grow (see Fig. \ref{fig:holes}).

\begin{figure}
\centering
\includegraphics[height=4cm]{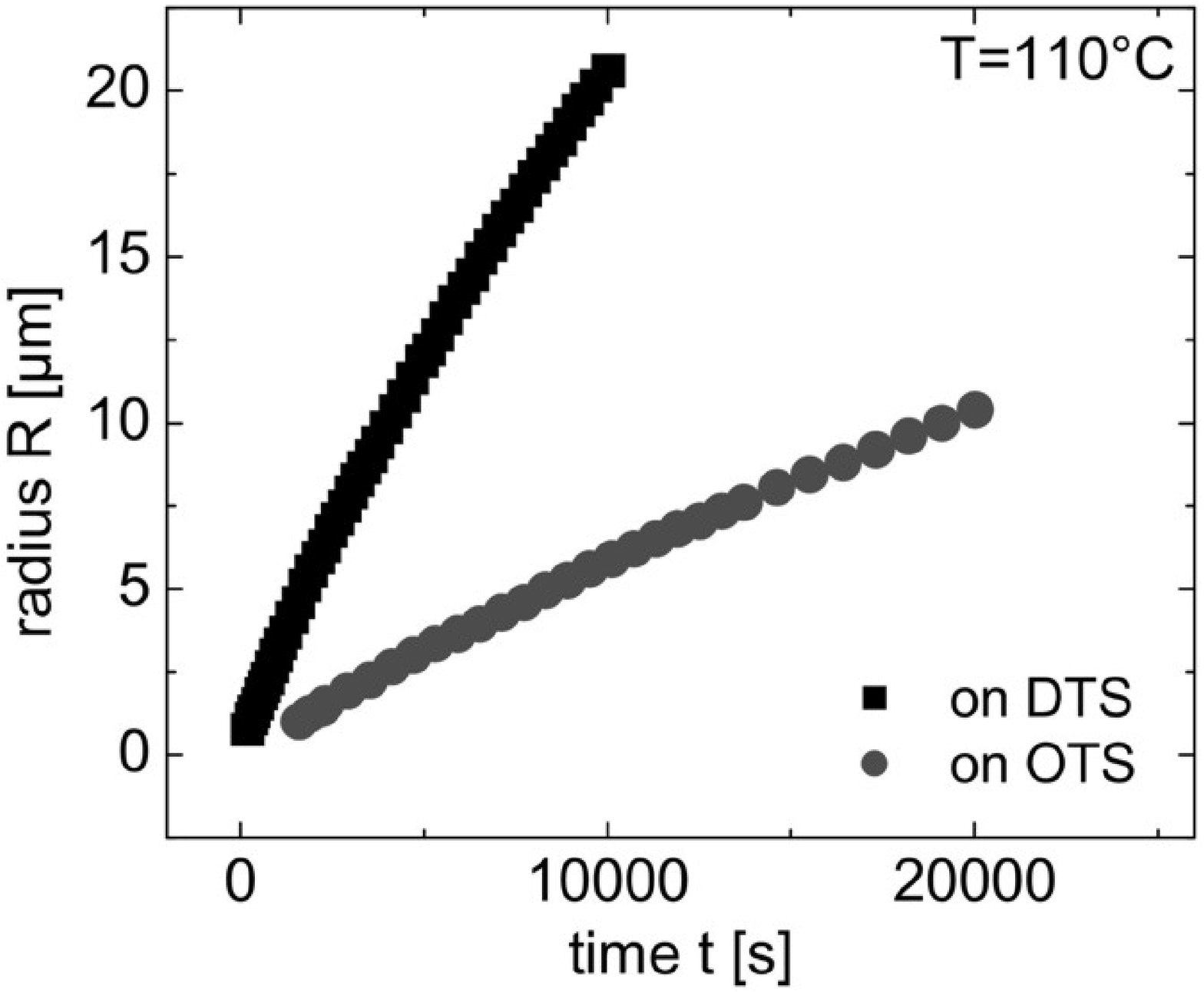}
\includegraphics[height=4cm]{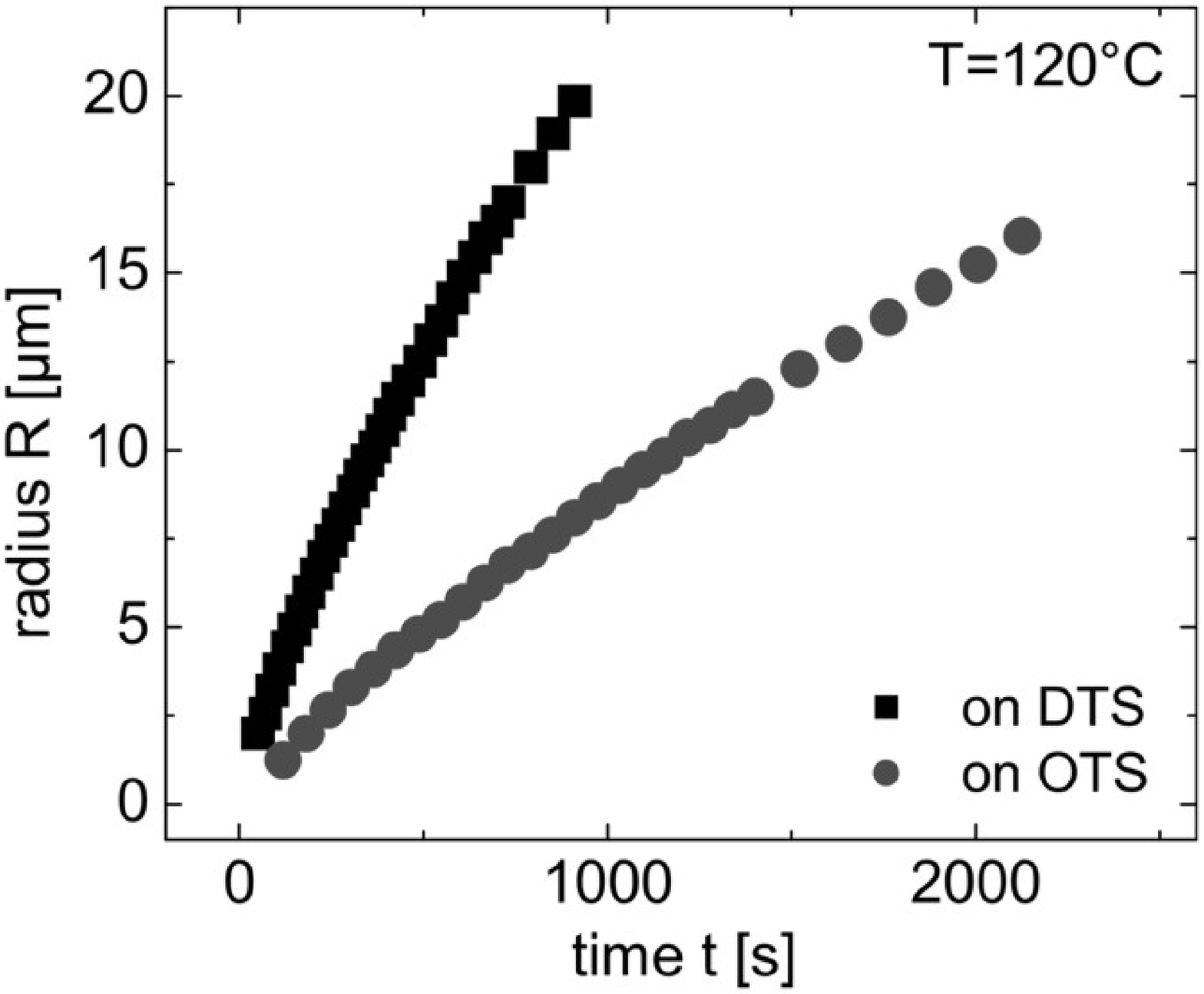}
\caption{Hole radius versus time on OTS and DTS at 110$^{\circ}$C
and 120$^{\circ}$C.}
\label{fig:r_t}       
\end{figure}

Holes in the PS film were imaged by optical microscopy captured by
an attached CCD camera, and hole radii were measured. As shown in
Fig. \ref{fig:r_t}, dewetting progresses much faster on DTS than
on OTS coated substrates.

We explain these results as follows: At the same temperature, the
liquids on both samples have exactly the same properties: the
viscosity as well as the surface tension do not depend on the
substrate underneath. Additionally, the contact angle of polystyrene
on both surfaces is the same within the experimental error.
Therefore, the spreading coefficient $S$, that is the driving force
of the dewetting process, is identical on both substrates. Hence,
the different dewetting velocities observed on OTS and DTS indicate
different energy dissipation pathways on these coatings. Viscous
friction within the liquid is expected to be identical for both
surfaces since we compare liquids of the same viscosity and with the
same dynamic contact angle \cite{Bro94}. The latter can be probed by
AFM scans in the vicinity of the three-phase contact line. In situ
imaging reveals that the dynamic contact angle stays constant at
$56(2)^{\circ}$ during hole growth. Consequently, friction at the
solid/liquid interface and therefore slippage must be different on
OTS and DTS. In the next section we show that not only the hole
growth dynamics but also the shape of rims is affected by the
underlying substrate.

\subsection{Rim Shapes}
\label{sec:2.3} 
\begin{figure}
\centering
\includegraphics[height=4cm]{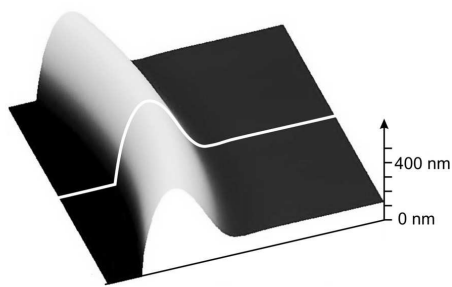}
\caption{AFM image of a rim on OTS: scan size
$10\times10\,\umu$m$^2$. The dewetting was performed at
130$^{\circ}$C. The white line represents a single scan line of the
rim perpendicular to the three-phase contact line.}
\label{fig:3dscan}       
\end{figure}

For the characterization of the shape of the rims
surrounding the holes, the samples were quenched to room temperature
after the holes have reached a diameter of $12\,\umu$m. AFM scans
were then taken in the glassy state of PS. An example is shown in
Fig. \ref{fig:3dscan}.

\begin{figure}
\centering
\includegraphics[height=4.5cm]{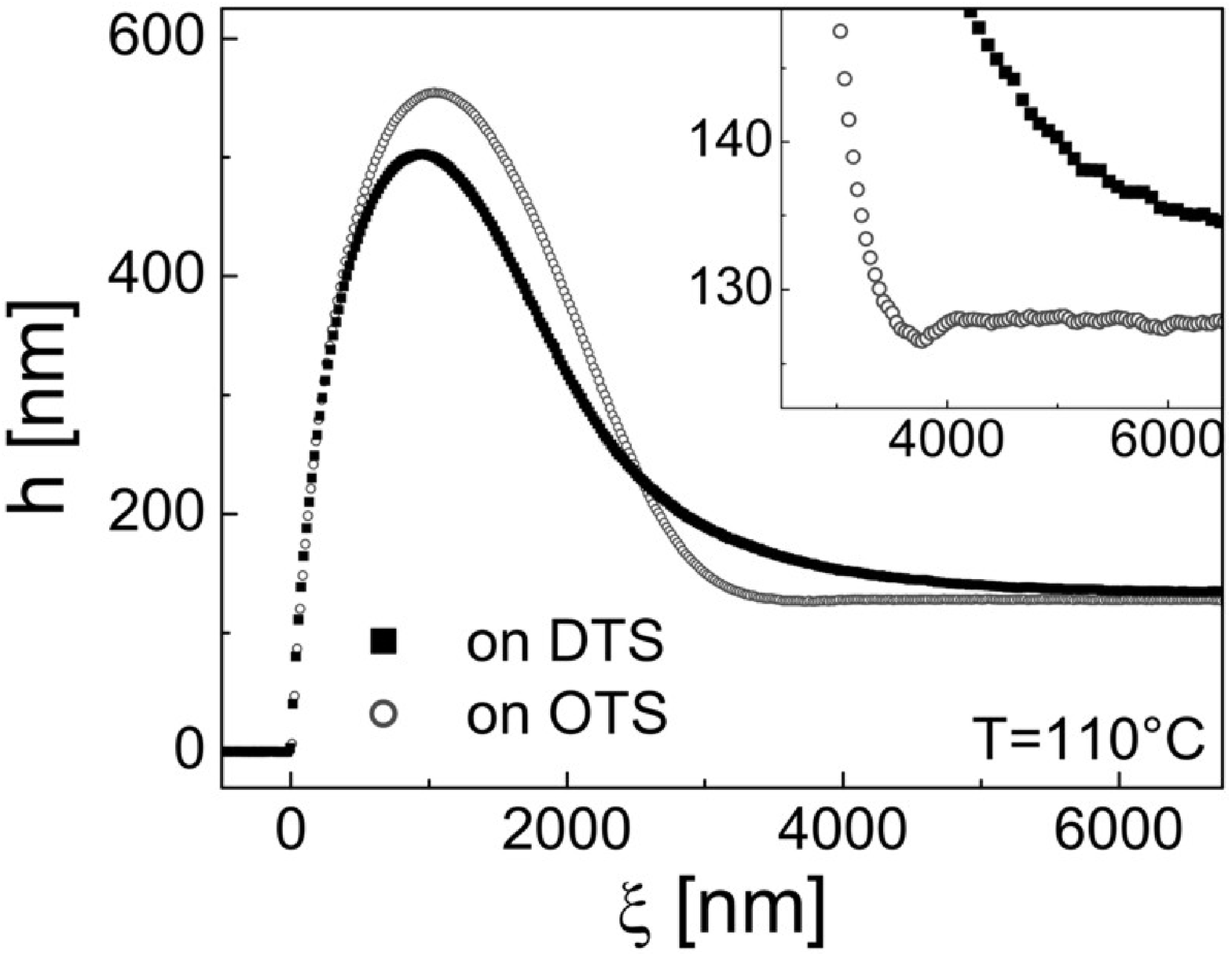}
\includegraphics[height=4.5cm]{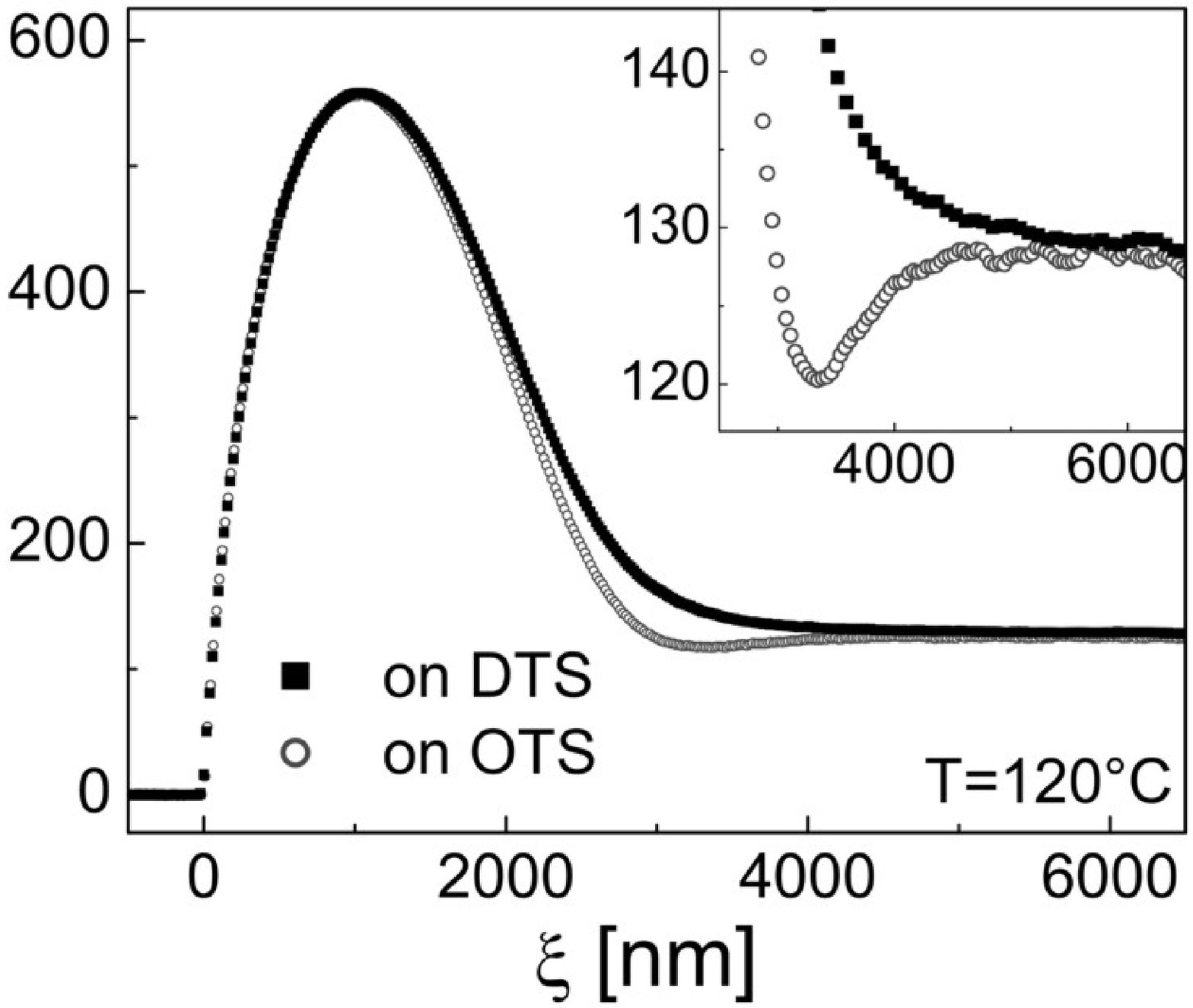}
\\
\includegraphics[height=4.5cm]{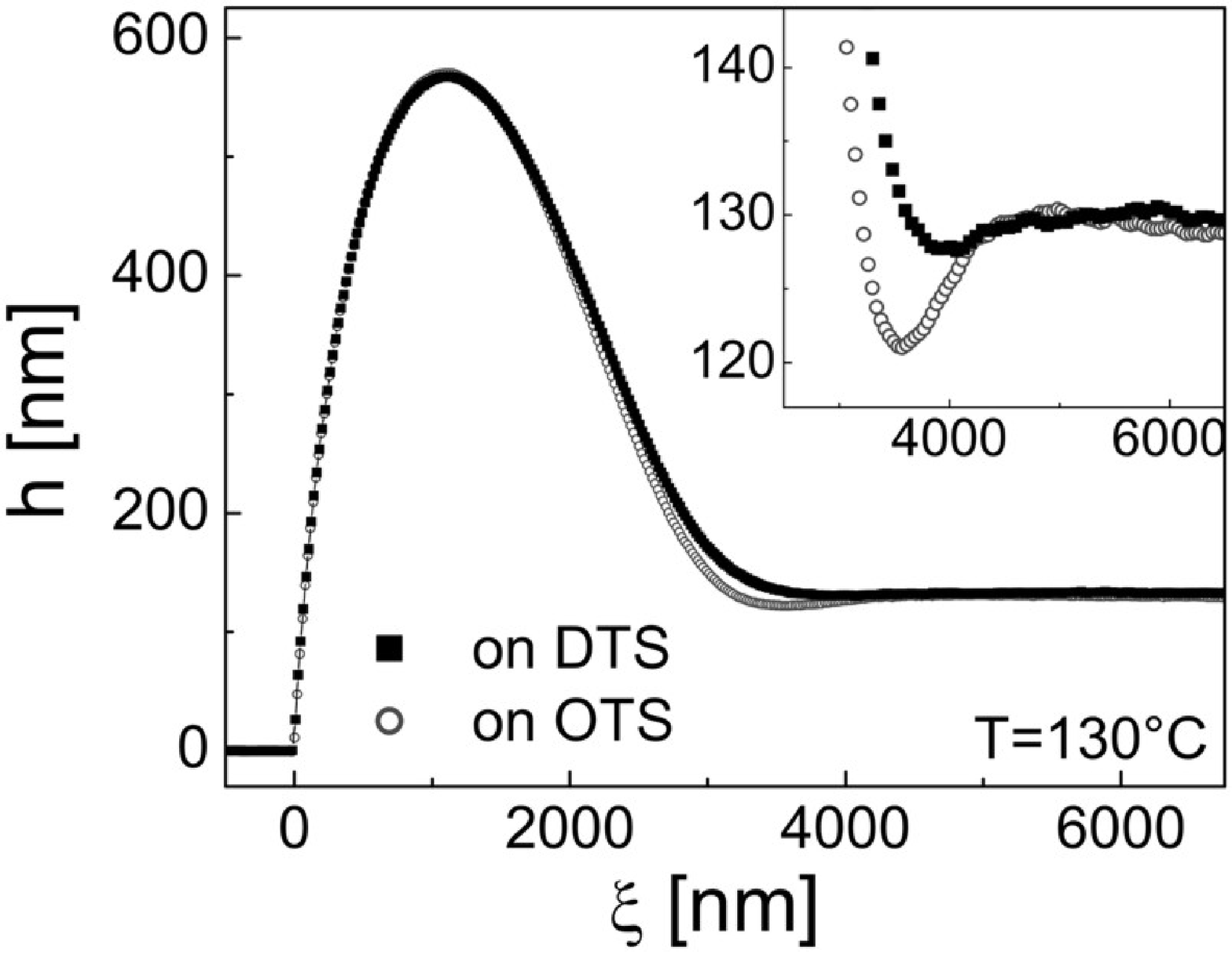}
\caption{Rim profiles on OTS and DTS at 110$^{\circ}$C,
120$^{\circ}$C and 130$^{\circ}$C.}
\label{fig:rims}       
\end{figure}

Fig. \ref{fig:rims} demonstrates that the type of substrate affects
the rim profile: On OTS covered substrates, the rim of the dewetting
PS film exhibits an oscillatory shape, whereas on DTS covered
surfaces, at the same temperature, a monotonically decaying function
is observed. The insets to Fig. \ref{fig:rims} shall clarify the
term "oscillatory rim shape" on OTS. Furthermore, Fig.
\ref{fig:rims} shows that temperature influences the shape of the
profile: the higher the temperature the more pronounced are the
oscillations on OTS and even on DTS, an oscillatory shape is
recorded for $T=130^{\circ}$C. Comparing the impact of the substrate
on the rim morphology to the hole growth experiments in the previous
section, we observe a correlation between the dewetting velocity and
the shape of corresponding rims: high dewetting velocities lead to
monotonic rim profiles while lower dewetting speeds tend to result
in oscillatory rim shapes. In the previous section we argued that
different sliding friction at the solid/liquid interface might be
responsible for the different dewetting speeds on OTS and DTS. In
the next section, the theoretical expectation of the influence of
slippage on the rim shapes will be explained. Moreover, the
experimental results are compared to theoretical predictions.

\section{Theoretical Models and Data
Analysis } \label{sec:3}

\subsection{Lubrication Model}
\label{sec:3.1} The theoretical description is based on a
lubrication model developed by M\"unch et al. \cite{Mue05} for
systems showing strong slip at the solid/liquid interface. Starting
point for this description are the Navier-Stokes equations in two
dimensions for a viscous, incompressible Newtonian liquid:

\begin{equation}\label{NavierStokes}
  -\nabla [p+\phi (h)] +\eta\nabla^2 \vec{u}= \rho (\partial_t \vec{u}+\vec{u}\cdot\nabla \vec{u}) \qquad \nabla \cdot \vec{u}=0,
\end{equation}

with pressure $p$, disjoining pressure $\phi (h)$, viscosity
$\eta$, density $\rho$ and the velocity $\vec{u}=(u,w)$ of the
liquid.


By applying the Navier slip boundary condition

\begin{equation}\label{Navier}
  b=\frac{u}{\partial_z u}|_{z=0}=\frac{\eta}{\kappa},
\end{equation}

where $\eta$ is the viscosity of the melt and $\kappa$ the
friction coefficient at the solid/liquid interface, and using a
lubrication approximation (which assumes that the typical length
scale on which variations occur is much larger in lateral
direction than in vertical direction) we get the following
equations of motion:

\begin{equation}\label{Bewegungsgleichung1}
u=\frac{2b}{\eta}\partial_x(2\eta h\partial_x u)+\frac{bh}{\eta}
\partial_x(\gamma \partial_x^2 h-\phi^\prime (h))
\end{equation}
\begin{equation}\label{Bewegungsgleichung2}
\partial_t h=-\partial_x(hu)
\end{equation}

Here, the slip length $b$ is assumed to be much larger than the
average film thickness $H$ (i.e., for $b/H\gg1$). In order to
perform a linear stability analysis of the flat film of thickness
$H$ we introduce a small perturbation $\delta h \ll H$ traveling
in the frame of the moving rim

\begin{equation}\label{Ansatz}
  h(x,t)=H+\delta h \exp(k\xi)
\end{equation}

with

\begin{equation}\label{Frame}
  \xi=x-s(t)
\end{equation}

where $s(t)$ is the position of the three-phase contact line. This
ansatz leads to the characteristic polynomial of third order in $k$:

\begin{equation}\label{LubTransition}
  (Hk)^3+4Ca(Hk)^2-Ca\frac{H}{b}=0
\end{equation}

One of the three solutions for $k$ is real and positive and
therefore does not connect the solution to the undisturbed film
for large $\xi$. Hence, this solution is not taken into account
any further. The remaining two solutions $k_1$ and $k_2$ are
either a pair of complex conjugate numbers with negative real part
or two real numbers $< 0$. These two different sets of solutions
correspond to two different morphologies of the moving rim, i.e.,
either an oscillatory ($k_1$ and $k_2$ are complex conjugate) or a
monotonically decaying ($k_1$ and $k_2$ are real) shape. These
morphologies are separated by a distinct transition as indicated
by the dashed line in Fig. \ref{fig:Cavsb2}. If both the capillary
number $Ca$ and the slip length $b$ are given for a certain film
thickness $H$, it is possible to qualitatively predict the shape
of the rim. This is in good agreement with our experimental
observations: On DTS we observe fast dewetting and expect large
slip lengths, and we indeed observe mainly monotonically decaying
rims; on OTS we find slower dewetting and expect comparatively
small slip lengths, and, again in agreement with the theoretical
prediction, we observe oscillatory profiles.

Furthermore, if two solutions $k_1$ and $k_2$ are known (or
extracted from experimentally obtained rim profiles), the
characteristic polynomial can by solved for the unknown slip
length $b$ and the capillary number $Ca$.  By that we obtain the
solution for $b$ and for $Ca$ from the strong-slip lubrication
model:

\begin{equation}\label{bLub}
    b_{lub}=\frac{1}{4H}\frac{k_1^2+k_1k_2+k_2^2}{k_1^2k_2^2}
\end{equation}

\begin{equation}\label{CaLub}
    Ca_{lub}=-\frac{H}{4}\frac{k_1^2+k_1k_2+k_2^2}{k_1+k_2}
\end{equation}

Besides this strong slip model, further lubrication models have
been developed \cite{Fet071,Mue05}. The corresponding weak slip
(i.e., for $b/H<1$) model always leads to complex solutions for
$k_1$ and $k_2$ and therefore predicts only oscillatory rim
shapes.

\subsection{Stokes Model}
\label{sec:3.2} As already mentioned, the former introduced
lubrication model is just valid for large slip lengths compared to
film thickness. To handle smaller slip lengths it is necessary to
use a more recently developed model \cite{Fet071} which is a
third-order Taylor expansion of the characteristic equation gained
by linear stability analysis of a flat film using the full Stokes
equations:

\begin{equation}\label{StokesTransition}
  (1+\frac{H}{3b})(Hk)^3+4Ca(1+\frac{H}{2b})(Hk)^2-Ca\frac{H}{b}=0
\end{equation}

\begin{figure}
\centering
\includegraphics[height=6cm]{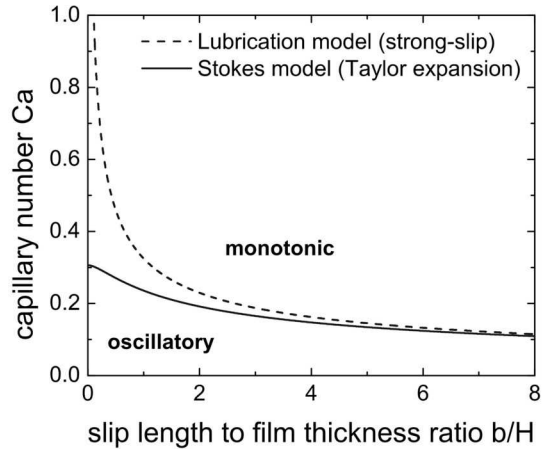}
\caption{Comparison of the theoretically expected transition from
oscillatory profiles to monotonic ones for the strong-slip
lubrication model (dashed line) and the third-order Taylor expanded
Stokes model (solid line).}
\label{fig:Cavsb2}       
\end{figure}

This equation again predicts a transition between oscillatory and
monotonically decaying rims. In Fig. \ref{fig:Cavsb2}, the
respective transition line (solid line) is compared to the one of
the strong-slip lubrication model (dashed line). As expected,
significant deviations between the two models occur for moderate and
weak slippage, while good agreement is given for strong slippage. As
shown in \cite{Fet071}, the transition line for the Taylor expansion
is, even for small slip lengths, quite close to the one predicted by
the full Stokes equations. Solving (\ref{StokesTransition}) for $b$
and $Ca$ leads to additional contributions compared to the
strong-slip lubrication model, (\ref{bLub}) and (\ref{CaLub}):

\begin{equation}\label{bStokes}
    b_{Taylor}=\frac{1}{4H}\frac{k_1^2+k_1k_2+k_2^2}{k_1^2k_2^2}-\frac{H}{2}
\end{equation}

\begin{equation}\label{CaStokes}
    Ca_{Taylor}=-\frac{H}{4}\frac{k_1^2+k_1k_2+k_2^2}{k_1+k_2}+\frac{H^3}{6}\frac{k_1^2k_2^2}{k_1+k_2}
\end{equation}

Note that the slip length obtained from the third-order Taylor
expansion of the characteristic equation of the full Stokes model
equals the slip length obtained from the strong-slip lubrication
model minus a shift of half of the film thickness.

We like to point out that, in contrast to previous publications
\cite{Fet05,Fet06}, all data sets presented in the following
sections were evaluated by applying the third-order Taylor expanded
Stokes model, i.e., (\ref{bStokes}) and (\ref{CaStokes}). This
enables us to exclude deviations that occur from the strong-slip
lubrication model, which is not valid for smaller slip lengths.

\subsection{Method to Extract Slip Length and Capillary Number}
\label{sec:3.3}

\begin{figure}
\centering
\includegraphics[height=6cm]{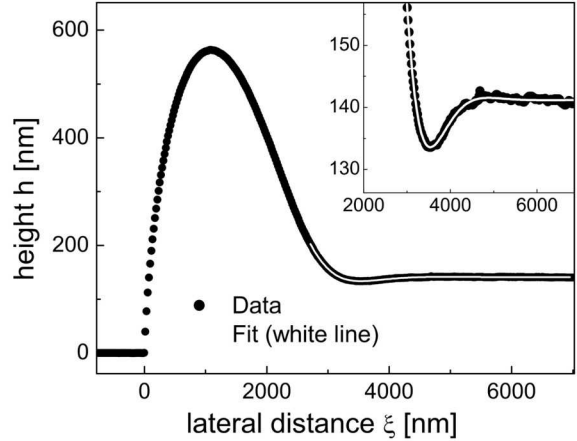}
\caption{AFM cross-section of a rim profile of a hole (about
$12\,\umu$m radius) in a PS(13.7k) film on OTS dewetted at
120$^{\circ}$C and corresponding fit.}
\label{fig:fit}       
\end{figure}

To fit the profile of a rim obtained by AFM we choose data points on
the "wet" side of the rim ranging from about $110\,\%$ of the film
thickness to the film thickness itself (see Fig. \ref{fig:fit}).
Oscillatory profiles are fitted by a damped oscillation as described
by (\ref{OsziFit}) with $\delta h_0$, $k_r$, $k_i$ and $\phi$ as fit
parameters:

\begin{equation}\label{OsziFit}
  \delta h_{osci} =\delta h_0\exp(k_r \xi)\cos(k_i\xi+\phi)
\end{equation}

$k_r$ is the real part and $k_i$ the imaginary part of $k_1$ and
$k_2$:

\begin{equation}\label{kOszi}
  k_{1,2}=k_r\pm ik_i
\end{equation}

In case of monotonically decaying rims we deal with a superposition
of two exponential decays given by (\ref{MonoFit}):

\begin{equation}\label{MonoFit}
  \delta h_{mono} =\delta h_1\exp(k_1 \xi)+\delta h_2\exp(k_2\xi)
\end{equation}

Here we obtain the amplitudes $\delta h_{1,2}$ and the decay lengths
$k_{1,2}$ as fit parameters. In some cases, the fitting procedure
may not be able to distinguish between two decaying exponentials,
leading to two identical decay lengths. In that case, an independent
calculation of $b$ and $Ca$ is not possible. However, if the
capillary number $Ca$ is known (for example if the viscosity is
known and the capillary number was calculated via (\ref{Capillary
number})), one decay length is sufficient to extract the slip
length.

\begin{figure}
\centering
\includegraphics[height=6cm]{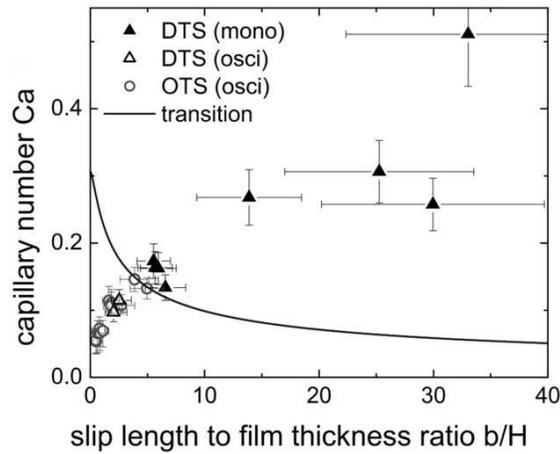}
\caption{Capillary number $Ca$ versus the ratio of slip length $b$
and film thickness $H$ on OTS and DTS obtained from the third-order
Taylor expanded Stokes model.}
\label{fig:Cavsb}       
\end{figure}

From the parameters $k_1$ and $k_2$ gained by fitting the respective
function (\ref{OsziFit}) or (\ref{MonoFit}) to the rim profiles, we
determined slip lengths via (\ref{bStokes}) and capillary numbers
via (\ref{CaStokes}) for the PS films on our substrates at different
temperatures, i.e., for different melt viscosities. As shown in Fig.
\ref{fig:Cavsb}, the capillary number clearly increases non-linearly
with the slip length. In the transition region from oscillatory
profiles to monotonic decaying rims, we have to deal with the fact
that profiles with just one clear local minimum, but no local
maximum in between the minimum and the undisturbed film, can be
fitted by (\ref{MonoFit}) as well as by (\ref{OsziFit}). In the
first case, this corresponds to one of the amplitudes $\delta
h_{1,2}$ being negative. We emphasize that exclusively
(\ref{OsziFit}) leads to slip lengths that do not depend on the hole
radius $R$, cf. the recent study \cite{Fet071}. This criterion
enables us to justify the choice of (\ref{OsziFit}) as the
appropriate fitting function in this region.

\subsection{Experimental Tests}
\label{sec:3.4}

\begin{figure}
\centering
\includegraphics[height=6cm]{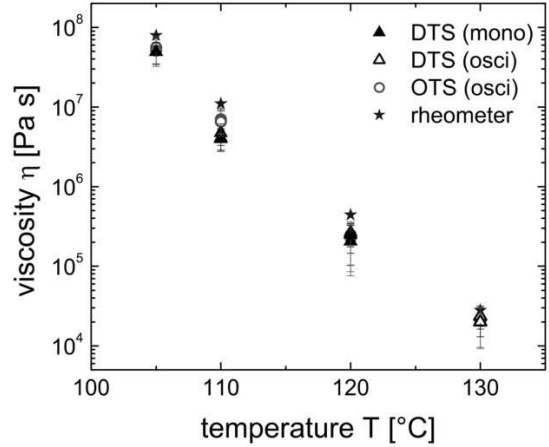}
\caption{Viscosities obtained from rim profile analysis and the
third-order Taylor expanded Stokes model compared to viscosimetry
data.}
\label{fig:etat}       
\end{figure}

The independent extraction of capillary number $Ca$ and slip length
$b$ from the rim profiles allows us to check the consistency of the
applied model. Comparing for instance the viscosities gained by rim
shape analysis (via the calculated capillary numbers, the dewetting
velocities $\dot{s}$ from hole growth experiments and
(\ref{Capillary number})) to independent viscosimetry measurements
shows excellent agreement for both types of substrates, cf. Fig.
\ref{fig:etat}.

Furthermore we evaluated certain rim shapes of the same hole but
at different hole radii, cf. \cite{Fet06}. Although the rim grows
in size during dewetting, the extracted slip length stays constant
within the range of our variation of rim size. The extracted
capillary number, however, decreases for increasing hole radius.
On the other hand, dewetting slows down with increasing rim size,
cf. Fig. \ref{fig:r_t}. The ratio of the measured dewetting
velocities $\dot{s}$ and the extracted capillary numbers gives
viscosity data independent of the hole size, which again is in
agreement with the expectation. To conclude, all these tests
underline the applicability of the former introduced models to our
system.

\section{Results and Discussion}
\label{sec:4} Analysis of rim shapes for different temperatures
above the glass transition temperature gives slip lengths ranging
from less than 100 nm up to about 5 microns.

\begin{figure}
\centering
\includegraphics[height=6cm]{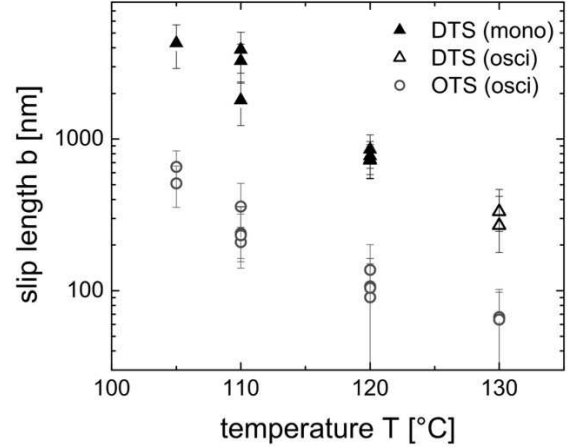}
\caption{Slip length $b$ on OTS and DTS versus the dewetting
temperature. Rim profile analysis was performed by applying the
third-order Taylor expansion of the Stokes model on holes of
$12\,\umu$m radius in a 130 nm PS(13.7k) film.}
\label{fig:bvsT}       
\end{figure}

\begin{figure}
\centering
\includegraphics[height=6cm]{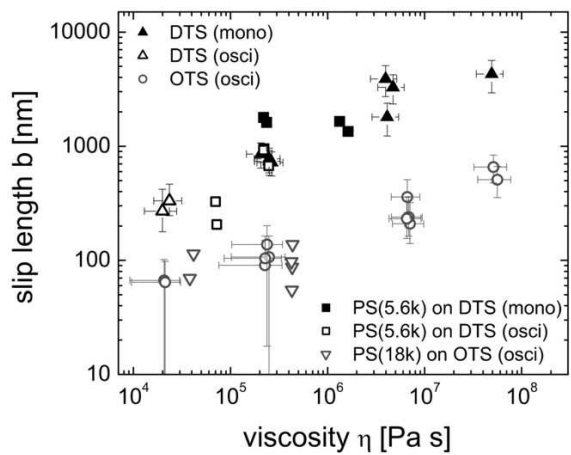}
\caption{Slip length data from Fig. \ref{fig:bvsT} versus the
respective film viscosity in logarithmic scale.}
\label{fig:beta}       
\end{figure}

As shown in Fig. \ref{fig:bvsT}, the slip length on DTS is about one
order of magnitude larger than on OTS. On both substrates, the slip
length decreases for higher temperatures. These results are in good
agreement with slip lengths obtained by hole growth analysis studies
\cite{Fet072}. Plotting the slip length versus melt viscosity,
determined by rim analysis, shows non-linear behavior (see
Fig.~\ref{fig:beta}). This is at variance with the Navier slip
condition (\ref{Navier}), where a linear dependency of slip length
on viscosity is expected. To evaluate this discrepancy, experiments
were performed with polymer melts of different molecular weights
below the entanglement length. The results are shown in
Fig.~\ref{fig:beta}. Variation of molecular weight of the melt gives
data that fall on a master curve for each substrate. This means that
friction has to be stronger for higher viscosities. This fact is
possibly a valuable hint to the molecular mechanism of friction and
slippage at the solid/liquid interface and will be further discussed
in the following.

A scenario described in literature \cite{Mig93,Herv03,Leg03} for
similar systems concerns melt chains penetrating between silane
molecules and adhering to the underlaying high-energy silicon
substrate. This may lead to polymer chains slipping over grafted
polymer chains of the same polymer. For this situation it is
reported that flow is expected be strongly dependent on shear rate
and grafting density. The density of adhered chains is assumed to be
dependent on the perfectness of the monolayer of the silane
molecules. From the surface energies we obtained for OTS and DTS, we
might expect OTS to build up a denser layer compared to DTS. Higher
shear rates obtained for larger dewetting rates i.e. temperatures
may lead to a dynamic desorption of adhered chains and, thus, change
friction at the interface. Molecular dynamics simulations of flowing
polymer chains interacting with grafted chains of the same polymer
by Pastorino et al. \cite{Pas06} have shown that interpenetration
and subsequent anchoring as well as the grafting density influences
slippage. They observe pronounced slippage for lower density of
anchored chains.

Our further studies concerning the molecular mechanism of slippage
focus on X-ray and neutron reflectivity experiments of (deuterated)
polystyrene melts on Si substrates covered with self-assembled
monolayers. These techniques are known to be very sensitive on
changes at the solid/liquid interface. First X-ray reflectivity
measurements on bare silanized wafers show that the silane molecules
form very dense monolayers and stand upright on top of the
underlying Si substrate \cite{Mag}. Therefore, the previously
discussed explanation of interdigitation does not seem to be very
likely. Other explanations assuming structural changes of the
substrate, such as bending or tilting of silane molecules, producing
a temperature-dependent slip length also seem not to be a major
issue in the temperature range of our experiments, though we can not
absolutely exclude them. Further, roughness has been shown in
miscellaneous studies to influence slippage \cite{Leg03,Cot03}.
However, due to the fact that our surfaces are extremely smooth, cf.
Sect. \ref{sec:2.1}, the influence of roughness might be safely
excluded. Other parameters such as the polarizability of the liquid
have been shown to influence slippage dramatically. Cho et al.
\cite{Cho04} observed lower slip length for higher polarizable
liquid molecules. In addition, the shape of the liquid molecules can
also be relevant for differences in slip lengths \cite{Sch05}. Using
the same liquid, the latter two aspects can not be responsible for
the different slip length on OTS and DTS.

The most probable scenario concerning the origin of the huge slip
lengths is the formation of a so-called lubrication layer, i.e., a
liquid layer of reduced viscosity close to the substrate, that may
build up due to migration of low molecular-weight species to the
solid/liquid interface or an alignment of liquid molecules at the
interface. This slip plane could cause large apparent slip lengths.
Systematic ellipsometry measurements on spots where a film front has
passed the substrate and potentially left a remaining liquid layer,
may corroborate this point of view. If lubrication layers are
important, then polydispersity of the liquid should play a crucial
role on slippage. Dewetting experiments dealing with mixtures or
double-layers of polymers with different chain lengths are planned.
At last, this argument cannot yet afford explaining the difference
in slippage on OTS and DTS of about one order of magnitude.
Nevertheless, we can speculate that the formation of this
lubrication layer may depend on the interfacial energies of PS and
silane brushes, $\gamma_{\mathrm{PS/OTS}}$ and
$\gamma_{\mathrm{PS/DTS}}$, which are known to be slightly
different, cf. Sect. \ref{sec:2.1}.

\section{Conclusion}
\label{sec:5} We could show that slippage may strongly affect
profiles of moving rims surrounding growing holes in thin liquid
films. Furthermore, the connection of fitting parameters concerning
the shape of the rims and system parameters as slip length $b$ and
capillary number $Ca$ via a theoretical model allows us to quantify
slippage. We observe a slip length which is about one order of
magnitude larger on DTS than on OTS and which decreases with
increasing temperature. Moreover, we record experimentally a
non-linear dependency of the viscosity on the slip length, which is
at variance with the expectation due to the Navier-slip condition.
Friction at the solid/liquid interface has to be enhanced for higher
viscosity of the melt. From variation of the molecular weight we
know that temperature can be assumed to be an indirect parameter in
our system influencing slippage via the direct parameter viscosity.
To conclude, we have presented a method to extract slip lengths from
the analysis of rim profiles of dewetting polymer films. This method
is a powerful tool to characterize slippage of dewetting liquid
films. Further studies will focus on the variation of substrate and
liquid properties. Viscoelastic properties of the liquid, i.e., high
molecular-weight polymer chains above the entanglement length, may
be topic of future research.

\section{Acknowledgment}
\label{sec:6} This work was financially supported by DFG grants JA
905/3 and MU 1626/5 within the priority program SPP 1164, the
European Graduate School GRK 532 and the Graduate School 1276. We
acknowledge the generous support of Si wafers from Siltronic AG,
Burghausen, Germany.

%
%
%
%
%

%
%



\end{document}